\documentclass[aps,pre,twocolumn,floatfix]{revtex4}
\usepackage[pdftex]{color}
\usepackage{latexsym,hyperref,amssymb,amsmath,gensymb}
\usepackage{graphicx,dcolumn,bm}
\def\geqsim{\lower.73ex\hbox{$\sim$}\llap{\raise.4ex\hbox{$>$}}$\,$}
\def\leqsim{\lower.73ex\hbox{$\sim$}\llap{\raise.4ex\hbox{$<$}}$\,$}
\newcommand{\half}{\frac{1}{2}}
\newcommand{\thalf}{{\textstyle{\frac{1}{2}}}}

\newcommand{\bos}{\boldsymbol}
\newcommand{\tbf}{\textbf}
\newcommand{\tit}{\textit}

\newcommand{\mbf}{\mathbf}

\newcommand{\beq}{\begin{equation}}
\newcommand{\eeq}{\end{equation}}
\newcommand{\bea}{\begin{eqnarray}}
\newcommand{\eea}{\end{eqnarray}}
\newcommand{\barr}{\begin{array}}
\newcommand{\earr}{\end{array}}
\newcommand{\bean}{\begin{eqnarray*}}
\newcommand{\eean}{\end{eqnarray*}}
\newcommand{\bei}{\begin{itemize}}
\newcommand{\eei}{\end{itemize}}
\newcommand{\ben}{\begin{enumeration}}
\newcommand{\een}{\end{enumeration}}
\newcommand{\nn}{\nonumber}
\newcommand{\tsc}{\textsc}

\newcommand{\la}{\langle}
\newcommand{\ra}{\rangle}
\newcommand{\lt}{\left}
\newcommand{\rt}{\right}
\newcommand{\ep}{\epsilon}
\definecolor{navyblue}{rgb}{.05,0,.55}
\newcommand{\tcr}[1]{\textcolor{red}{#1}}

\begin{document}
\begin{widetext}
\begin{flushleft}
{\Large{\textbf{\textsf{Molecular Electroporation and the Transduction of Oligoarginines}}}}
\end{flushleft}

\begin{flushleft}
\textsf{Kevin Cahill\hfill\today\\cahill@unm.edu}\\
\textsf{Biophysics Group,
Department of Physics \& Astronomy,
University of New Mexico, Albuquerque, NM 87131}
\end{flushleft}

\pagestyle{myheadings}
\markright{Electroporation and Transduction}

\begin{abstract}\noindent
\textsf{ABSTRACT \quad
Certain short polycations, such as TAT and polyarginine, 
rapidly pass through the plasma membranes
of mammalian cells 
by an unknown mechanism called transduction
as well as by endocytosis and macropinocytosis.
These cell-penetrating peptides (CPPs)
promise to be medically useful
when fused to biologically active peptides.
I offer a simple model in which 
one or more CPPs and the phosphatidylserines
of the inner leaflet form a kind of capacitor
with a voltage in excess of 180 mV,
high enough to create a 
molecular electropore.
The model is consistent with
an empirical upper limit 
on the cargo peptide of 
40--60 amino acids and with experimental data 
on how the transduction of a polyarginine-fluorophore
into mouse C\(_2\)C\(_{12}\) myoblasts
depends on the number of 
arginines in the CPP and on the CPP concentration.
The model makes three testable predictions.}
\end{abstract}

\maketitle
\end{widetext}
\section{Cell-Penetrating Peptides
\label{Cell-Penetrating Peptides}}
In 1988, two groups~\citep{Loewenstein1988,Pabo1988} 
working on HIV
reported that the \tcr{t}rans-\tcr{a}ctivating 
\tcr{t}ranscriptional activator
(\tcr{TAT}) of HIV-1 can cross cell membranes.
The engine driving this 86-aa 
cell-penetrating peptide (CPP)
is its residues 48--57 
\textsc{grkkrrqrrr}
which carry a charge of +8\(e\)\@.
Other CPPs soon were found.
Antp (aka Penetratin, PEN) is residues 43--58
\(\textsc{rqikiwfqnrrmkwkk}\)
of Antennapedia, a homeodomain
of the fly; it carries a charge of +7\(e\)\@.
The polyarginine (Arg)\(_n\) carries charge
\(+ n e \), where often \(n = 7, 8\), or 9\@.
Other CPPs have been discovered
(VP22) or synthesized (transportan)\@.
The structural protein VP22
of the tegument of herpes simplex virus type 1 (HSV-1)
has charge +15\(e\)\@.
Transportan
\(\textsc{gwtln}\textsc{sagyllg}\)-\(\textsc{k}\)-\(\textsc{inlkala}\textsc{alakk}\textsc{il}\)-amide
is a chimeric peptide constructed from 
the 12 N-terminal residues of galanin 
in the N-terminus with the 14-residue
sequence of mastoparan  
and a connecting lysine~\citep{Lindberg2001}\@.
With its terminal amide group,
its charge is +5\(e\)\@.
\par
These and other short, positively
charged peptides can penetrate 
the plasma membranes of live cells
and can tow along with them cargoes
that greatly exceed the 600 Da restriction barrier.
They are promising therapeutic 
tools when towing cleverly chosen peptide cargoes 
of from 8 to 33 amino acids~\citep{Tsien2004,Dowdy2005,Pugh2002,Kaelin1999,Fong2003,Cohen2002,Robbins2001,Datta2001,Hosotani2002,Hsieh2006,SnyderDowdy2004,Morano2006,Tuennemann2007}\@.
\par
Many early experiments on CPPs 
were wrong because the
cells were fixed or insufficiently
washed.  Even careful experiments 
sometimes have yielded inconsistent 
results---in part
because fluorescence varies with the 
(sub)cellular conditions and 
the fluorophores~\citep{Seelig2007}.
\par
Yet some clarity is emerging:
TAT carries cargoes across 
cell membranes with high efficiency
by at least two functionally distinct mechanisms
according to whether the cargo is big
or small~\citep{Cardoso2006}\@.
Big cargoes, such as proteins
or quantum dots, enter via 
caveolae endocytosis
and macropinocytosis~\citep{Dowdy2004,Brock2007},
and relatively few escape the 
cytoplasmic vesicles in which they then
are trapped~\citep{Cardoso2006}\@.
\par
Small cargoes, such as peptides 
of fewer than 30--40 amino acids, 
enter both slowly
by endocytosis and rapidly by transduction
with direct access to the cytosol,
an unknown mechanism
that uses the membrane 
potential~\citep{Cardoso2006,Prochiantz2000,Rezsohazy2003,Wender2005,Shen2005}\@.
Peptides fused to TAT
enter cells
within seconds~\citep{Seelig2005}\@.
\par
It remains unclear how big cargoes
aided by several CPPs enter cells~\citep{Patel2007}\@.
For instance,
superparamegnetic nanoparticles
encased in aminated dextran and
attached to 45 tat peptides are
thought to enter cells by
adsorptive endocytosis\citep{Fawell1994,Nagahara1998,Bulte2006} 
but they do enter slowly at 
4\(^\circ\) C~\citep{Garden2006}\@.
\par
This paper is exclusively about 
how polycationic cell-penetrating peptides, 
specifically oligoarginines,
transduce small cargoes directly
into the cytosol.
Sec.~\ref{Mammalian Plasma Membranes}
recalls some basic facts about
plasma membranes, and
Sec.~\ref{The Problem} explains why
ions do not normally pass through
plasma membranes.
Sec.~\ref{The Model}
describes a simple model of the transduction
of CPPs in which electroporation
and phosphatidylserine
play key roles.  
In this model, one or more positively charged CPPs
on the outer leaflet and the negatively charged
PSs under it on the inner leaflet form a kind 
of capacitor, which enhances the membrane
potential to a voltage in excess of 180 mV,
which is sufficient to create an electropore.
Sec.~\ref{Comparison with Experiment} shows that
the model is consistent with an 
empirical upper limit on the cargo
of 40--60 amino acids and
with measurements made by
T\"{u}nnemann \textit{et al.}~\citep{Cardoso2008}
on the fraction of mouse myoblasts 
transduced by polyarginines carrying fluorophores
of 400 Da.
Sec.~\ref{Three Tests of the Model}
tells how to test three predictions of the model.
The paper ends with a short summary
in Sec.~\ref{Summary}\@.

\section{Mammalian Plasma Membranes
\label{Mammalian Plasma Membranes}}
The plasma membrane of a mammalian cell
is a lipid bilayer that is 4 or 5 nm thick.
Of the four main phospholipids in it,
three---phosphatidylethanolamine (PE),
phosphatidylcholine (PC), and
sphingomyelin (SM)---are neutral,
and one, phosphatidylserine (PS), 
is negatively charged\@.
In live cells, PE and PS 
are mostly in the cytosolic layer,
and PC and SM in the 
outer layer~\citep{Zwaal1999,MBoC4587}\@.  
Aminophospholipid translocase (flippase) moves 
PE and PS 
to the inner layer;
floppase slowly moves all phospholipids
to the outer layer~\citep{Zwaal1999}\@.
\par
Incidentally, the surfaces of bacteria are different.
The cell wall of a Gram-positive bacterium 
(\textit{e.g., Streptococcus} or 
\textit{Staphylococcus})
is covered with negatively charged teichoic acids;
the outer leaflet of the outer membrane
of a Gram-negative bacterium 
(\textit{e.g., E.~coli} or \textit{Salmonellum})
is tiled
by negatively charged lipopolysaccharides (LPS)
held together by divalent 
cations~\citep{Vaara1992,Brogden2005}\@.
\par
Glycolipids make up about 5\%
of the lipid molecules of the outer layer
of a mammalian plasma membrane
where they may form lipid rafts.
Their hydrocarbon tails normally
are saturated.  Instead of a modified
phosphate group, they are decorated 
with galactose, glucose, 
GalNAc = N-acetylgalactosamine,
and other sugars.  The most complex
glycolipids---the gangliosides---have 
negatively charged 
sialic-acid (NANA) groups\@.
Incidentally, cholera toxin binds to and enters cells 
that display the G\(_{M1}\) 
ganglioside.~\citep{MBoC4587}
\par
A living cell maintains an
electrostatic potential of 
between 20 and 120 mV
across its plasma membrane.
The electric field \(E\) within
the membrane points into the cell
and is huge, about 15 mV/nm or
\(1.5 \times 10^7\) V/m
if the potential difference 
is 60 mV across a membrane of 4 nm.
Conventionally, one reports membrane
potentials as the electric potential
inside the cell minus that outside,
so that here \(\Delta V = - 60\) mV\@.
Near but outside the membrane,
this electric field falls-off
exponentially \(E(r) = E \, \exp(-r/D_\ell)\)
with the ratio of the distance \(r\)
from the membrane to the Debye length \(D_\ell\),
which is of the order of a nanometer\@.
The rapid entry of TAT fused to peptides is frustrated
only by agents that destroy the
electric field \(E\)~\citep{Cardoso2006},
which applies a force \(qE\)
to a CPP of charge \(q\)\@.
\par
Most of the phospholipids 
of the outer leaflet of the plasma membrane
are neutral PCs \& SMs\@.
They vastly outnumber
the negatively charged gangliosides, 
which are a subset
of the glycolipids, which themselves amount only
to 5\% of the outer layer.
Imagine now that 
CPP-cargo molecules
are in the extra-cellular environment.
Many of them 
will be pinned down by the electric field
\(E(r)\) just outside the membrane,
their positively charged side-chains
interacting with the negative phosphate
groups of neutral
dipolar PC \& SM head groups~\citep{Wender2005}.
(Other CPP-cargo molecules 
will stick to negatively
charged gangliosides and  
to glycosaminoglycans (GAGs) 
attached to transmembrane proteoglycans (PGs);
these slowly will be endocytosed.
PGs with heparan-sulfate GAGs are needed
for TAT-protein endocytosis~\citep{Giacca2001}\@.
A more detailed analysis than that of this work
might model the effect of these anionic
matrix compounds upon transduction.)
It is crucial that the 
dipolar PC \& SM head groups 
are neutral and so do not
cancel or reduce the positive electric charge
of a CPP-cargo molecule.
The net positive charge of a CPP-cargo
molecule and the negatively charged
PSs under it on the inner leaflet
form a kind of capacitor.
This is the starting point for the model
described in Sec.~\ref{The Model}\@.

\section{The Problem
\label{The Problem}}

The dielectric constant 
\(\epsilon_{\ell} \approx 2\) 
of the hydrocarbons of a lipid bilayer
is much less than that of
water \(\epsilon_w \approx 80\)\@.  
Thus, the difference \(\Delta E_{w \to \ell}\) 
in the electrostatic
energy of an ion of charge \(q\)
and effective radius \(a\) in the bilayer
and in water~\citep{Parsegian1969} is
\beq
\Delta E_{w \to \ell} = \frac{q^2}{8 \pi \epsilon_0 a}
\lt( \frac{1}{\epsilon_{\ell}} 
-  \frac{1}{\epsilon_w} \rt)
\label {Delta E wl}
\eeq
or 3.5 eV 
if the ion's charge is that
of the proton and its
radius is \(a = 1\) \AA\@. 
This energy barrier is far larger
than the 0.06 eV gained
when a unit charge crosses
a 60 mV phospholipid bilayer.
Thus, an ion will not cross
a cell's plasma membrane
unless a transporter or a channel 
facilitates (and regulates) its passage.
\par
In the present model of CPP transduction,
the electrostatics
of the CPP-cargo complex and 
the role of PSs on the inner leaflet
play key roles.
\par
The electrostatics of 
a cationic polypeptide such
as TAT or polyarginine are
more complex than for an ion.  
I will model
the CPP and its cargo
in water as a sphere with its 
positive charges on its surface.
The density of a protein
of mass \(M\) kDa is estimated~\citep{Craievich2004}
to be
\beq
\rho(M) = \lt( 0.8491 + 0.0873 \, e^{-M/13} \rt)
\quad \mbox{kDa/nm}^3.
\label {rho(M)}
\eeq
A CPP-cargo complex would not be expected
to fold
as densely as a natural globular protein, 
and so for it the
estimate \(\rho(M)\) is something of an upper bound.
The radius \(r\) of a putative sphere 
consisting of \(M\) kDa of CPP and cargo
then would be 
\beq
r \gtrsim \lt(\frac{3}{4\pi} 
\frac{M}{\rho(M)}\rt)^{1/3}
\quad \mbox{nm}.
\label {r =}
\eeq
For instance,  
a CPP of \(N\)
arginines and a tiny fluorophore
cargo of 400 Da 
has a mass of \(M_N = 0.1562 N + 0.4\) kDa,
and so its radius would satisfy
\beq
r \gtrsim  \lt(\frac{3}{4\pi} 
\frac{M_N}
{\rho(M_N)}\rt)^{1/3} 
\quad \mbox{nm}
\label {r NR}
\eeq
or \(r = 0.75\) nm
for \(N = 8\) arginines.
The lower bounds on the radii for \(N = \) 5--12
are listed in column 2 of 
the Table~(\ref{table of energies})\@.
\par
For larger cargoes
of \(A = 50\)--100 amino acids
of 130 Da each, the lower bounds
on the radii
range from 1.25 to 1.59 nm.
(In what follows, \(A\) will represent
the number of amino acids in the cargo
or the mass of the cargo in Daltons
divided by 130 Da\@.)
Adding another 0.8 nm for the PC/SM
head groups would extend these lower bounds
on the radii
to 2.05--2.39 nm.  
\par
If the CPP-cargo molecule were
a charged conducting sphere
of radius \(r\) and charge \(q\),
then its electrostatic energy
in water would be
\beq
E(N,A,q,w) = 
\frac{q^2}{8\pi \epsilon_0 \epsilon_w r}.
\label {E(N,A,w)}
\eeq
This term neglects the
short-distance detail of
the electric field near
the \(q/e\) positive unit charges \(e\)
of the CPP-cargo molecule.  
So a short-distance correction term 
\beq
E_{sdc}(a,q,w) = 
\frac{qe}{8\pi \epsilon_0 \epsilon_w a}
\label {Ecw}
\eeq
proportional to \(q\) 
must be added to \( E(N,A,q,w) \)\@.
The short distance \(a\)
is a parameter,
which will turn out to
be a few \AA\ because
the term \(E_{sdc}\) 
is a correction to be added
to \(E(N,A,q,w)\) and not
the entire electrostatic energy.
\par
The electrostatic field of
the cell attracts the CPP-cargo molecule
to the surface of the cell.
While on the outer leaflet
of the plasma membrane,
the electrostatic energy
of the CPP-cargo molecule
and its short-distance correction
are no longer given by 
their values
in water 
Eqs.~(\ref{E(N,A,w)} \& \ref{Ecw})
but instead are those appropriate
to the interface between water and lipid
\beq
E(N,A,q,w\ell) = 
\frac{q^2}{8\pi \epsilon_0 \bar \epsilon r}.
\label {E(N,A,q,w)}
\eeq
and
\beq
E_{sdc}(a,q,w\ell) = 
\frac{qe}{8\pi \epsilon_0 \bar \epsilon a}
\label {Ecqw}
\eeq
where
\( \bar \epsilon \)
is the mean permittivity
\beq
\bar \epsilon = \half \lt( \epsilon_w
+ \epsilon_\ell \rt).
\label {bareps}
\eeq
\par
The CPP-cargo molecule 
enters the lipid bilayer
as a CPP-cargo-PC/SM complex
with the phosphate groups
of the PC and SM of the
outer leaflet bound to
the positively charged
guanidinium and amine groups
of the CPP~\citep{Wender2005}.  
The positive charges of the
phosphocholine groups of PC and
SM are about \(d =5\) \AA\  from
their phosphate groups~\citep{MBoC5.620}\@.
The binding of PC and SM
therefore approximately
increases the effective
radius of the charged
sphere to \(r_m \approx r + d\)\@.
The electrostatic energy
of this complex in the hydrocarbon
tails of the lipid bilayer then is
\beq
E(N,A,q,\ell) \approx
\frac{q^2}{8\pi \epsilon_0 \epsilon_{\ell} (r+d)}
\label {E(N,A,m)}
\eeq
apart from a short-distance correction factor
\beq
E_{sdc}(a,q,\ell) = 
\frac{qe}{8\pi \epsilon_0 \epsilon_{\ell} a}
\label {Ecl}
\eeq
similar to (\ref{Ecw})\@.
\par
Apart from correction terms,
the electrostatic energy penalty when the
CPP-cargo molecule enters the lipid bilayer
from water as a CPP-cargo-PC/SM
complex is the difference
\bea
\Delta E_{w, \ell}^0(N,A,q) 
& \approx & E(N,A,q,\ell) - E(N,A,q,w\ell) 
\label {dE NAq} \\
& \approx & \frac{q^2}
{8\pi \epsilon_0 \epsilon_{\ell} (r+d)}
\lt( 1 - \frac{r+d}{r}
\frac{\epsilon_{\ell}}{\bar \epsilon}
\rt). \nn
\eea
\par
Because the thickness \(\ell = 4.4\) nm
of the lipid bilayer
is at most a few times the diameter of the
CPP-cargo-PC/SM complex, we also
must include the 
Parsegian correction~\citep{Parsegian1969}
\beq
\Delta E_P = - 
\frac{q^2}{4\pi \epsilon_0 \epsilon_{\ell} \ell}
\ln\lt(\frac{2 \epsilon_w}{\epsilon_w + \epsilon_{\ell}}
\rt)
\label {E_P}
\eeq
which holds when a uniformly charged sphere
is inserted into the middle of a lipid layer
of thickness \(\ell\)\@.
The sum of the water-to-lipid energy (\ref{dE NAq})
and Parsegian's correction (\ref{E_P}) is
\beq
\Delta E_{w, \ell}(N,A,q) = 
\Delta E_{w, \ell}^0(N,A,q) + \Delta E_P .
\label {dEwlP}
\eeq
The energy \(\Delta E_{w, \ell}(N,A,q)\)
is listed in column~3 of 
Table~(\ref{table of energies})
for a CPP of \(N = \) 5--12 arginines towing
a fluorophore cargo of 400 Da
with \(d = 0.5\) nm.
\par
The short-distance correction terms 
augment this penalty by 
\bea
\Delta E_{sdc}(a,q) & = & E_{sdc}(a,q,\ell)
- E_{sdc}(a,q,w) \nn\\
& = & \frac{qe}{8\pi \epsilon_0 \epsilon_{\ell} a}
\lt( 1 - \frac{\epsilon_{\ell}}{\epsilon_w} \rt)
\label {dEsdc}
\eea
and do not require Parsegian's correction
because they are short-distance effects.
This short-distance correction \(\Delta E_{sdc}\)
is listed in column~4 of 
Table~(\ref{table of energies})
for CPPs of \(N = \) 5--12 arginines
and a representative value of 
\(a = 4.5\) \AA\ for the short-distance parameter.
\par
The net electrostatic energy penalty when the
CPP-cargo molecule enters the lipid bilayer
from water as a CPP-cargo-PC/SM
complex is then the sum of 
(\ref{dE NAq}, \ref{E_P}, \& \ref{dEsdc})
\beq
\Delta E_{w \to \ell} =
\Delta E_{w,\ell}^0 + \Delta E_P + \Delta E_{sdc}.
\label {full electrostatic cost}
\eeq
A CPP of 8 arginines carrying
a fluorophore of 400 Da (\(A = 3 \approx 400/130\))
has a radius
\(r\) of 0.75 nm, and with \(a = 4.5\) \AA,
the change (\ref{full electrostatic cost})
in its electrostatic energy
on going from water to lipid is 
\beq
\Delta E_{w \to \ell} \approx
16.9 \; \mbox{eV}.
\label {dE(9,15,9e)}
\eeq
This energy barrier is 35
times bigger than the energy
0.48 eV that
it gains by crossing
a potential difference of 60 mV\@.
So how and why does it cross?

\section{The Model
\label{The Model}}

\par

\begin{table}
\caption{\label{tab:table of energies}
The radius \(r\) of a CPP-cargo molecule
of \(N\) arginines and a cargo of 400 Da,
its change in electrostatic energy 
\(\Delta E_{w,\ell}\)
when transferred from water to hydrocarbon
(Eq.~\ref{dEwlP}), and
the short-distance correction \(\Delta E_{SDC}\)
(Eq.~\ref{dEsdc})\@.
Distances are in nm and energies in eV\@.}
\label {table of energies}
\begin{ruledtabular}
\begin{tabular}{||c|c|c|c||} 
\, \(N\) \quad & \(r\) \quad
& \(\Delta E_{w, \ell}\) \quad  
& \(\Delta E_{sdc}\) \quad \\ \hline 
5 & 0.67 & 4.61 & 3.90  \\ \hline
6 & 0.70 & 6.39 & 4.68  \\ \hline
7 & 0.73 & 8.41 & 5.46  \\ \hline
8 & 0.75 & 10.64 & 6.24 \\ \hline
9 & 0.78 & 13.06 & 7.02  \\ \hline
10 & 0.80 & 15.68 & 7.80  \\ \hline
11 & 0.82 & 18.48 & 8.58  \\ \hline
12 & 0.84 & 21.44 & 9.36  \\ \hline
\end{tabular}
\end{ruledtabular}
\end{table}

My answer is that one (or more) oligoarginines and
the phosphatidylserines (PSs) of the inner leaflet 
together with their counterions
form a kind of capacitor with an electric
field strong enough to form a reversible pore
in the plasma membrane.
The transmembrane potential is the sum
of three terms---the resting transmembrane
potential \(\Delta V_{cell}\) of the cell
in the absence of CPPs, the transmembrane
potential \(\Delta V_{CPP}\) due to an
oligoarginine, and the  transmembrane
potential \(\Delta V_{NaCl}\) due to
the counterions of the extracellular medium
\beq
\Delta V = \Delta V_{cell} + 
\Delta V_{CPP} + \Delta V_{NaCl}.
\label {dV3}
\eeq
The resting transmembrane
potential \(\Delta V_{cell}\) of the cell
varies between about 20 mV 
to more than 70 mV, depending upon
the type of cell.  Ideally, it is
measured experimentally.
\par
My model is based upon several considerations,
which I discuss in turn in this section.
The first subsection describes the basic facts
about electroporation.  The second subsection
presents the electric potential \(V\) due to a 
charge in the extracellular medium;
the derivation of that potential is in
an appendix.  This potential implies
that charges on opposite sides
of the lipid bilayer are effectively
decoupled, which simplifies the
subsequent analysis.  The third subsection 
describes a Monte Carlo
simulation of the response of the
phosphatidylserines of the inner leaflet
to an oligoarginine interacting
with the phosphate groups of the outer leaflet.
To a very good approximation,
the PSs are distributed uniformly
and randomly because they are nearly decoupled from
the oligoarginine.
The fourth subsection uses the potential
\(V\) to compute the contribution \(\Delta V_{CPP}\)
of an oligoarginine to the transmembrane potential.
The fifth subsection describes
a Monte Carlo simulation of the 
effect \(\Delta V_{NaCl}\)
of the sodium and chloride ions in the
extracellular medium upon the transmembrane potential.
extracellular medium.  
The section ends with a summary of the model.

\subsection{Electroporation
\label{Electroporation}}

\begin{figure}
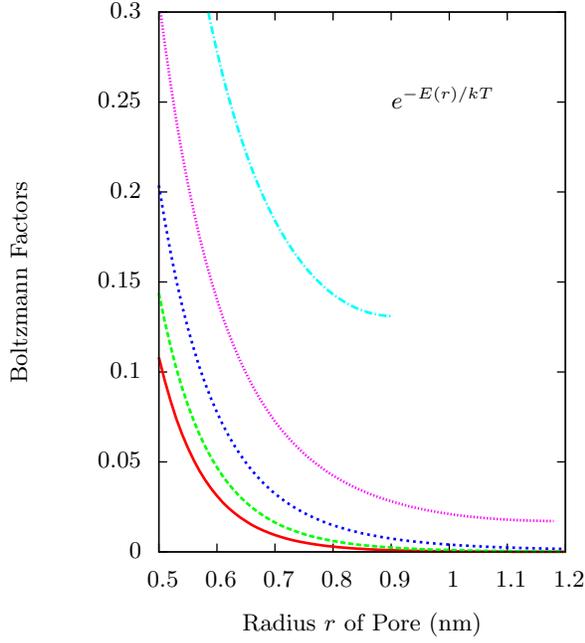

\centering
\input BolFac
\caption{The Boltzmann factor \(e^{-E(r)/kT}\)
(\(\times 100\))
for energy \(E(r)\) (Eq.(\ref{energy of pore}))
is plotted against the radius \(r\) of the pore
up to the critical radius \(r_c\) (Eq.(\ref{r critical}))
for the transmembrane voltage \(\Delta V = - 200\) (solid red),
\(- 250\) (dashes green), \(- 300\) (dots blue), 
\(- 350\) (dots magenta), and \(- 400\) mV (dash-dot cyan)\@.}
\label {BolFac}
\end{figure}

Electroporation is the formation
of pores in membranes by an electric field.
Depending on the duration of the
field and the type of cell,
an electric potential difference
across a cell's plasma membrane 
in excess of 150 to 200 mV will
create pores.  There are two main
components to the energy of a pore.
The first is the line energy 
\(2 \pi r \gamma \)
due to the linear tension \(\gamma\),
which is of the order of 
\(10^{-11}\) J/m\@.  The second
is the electrical energy
\( - 0.5 \Delta C \pi r^2 (\Delta V)^2 \)
in which \(\Delta V\) is the voltage
across the membrane and
\(
\Delta C = C_w - C_\ell 
\)
is the difference between the specific
capacity per unit area \(C_w = \ep_w \ep_0/t\)
of the water-filled pore
and that \(C_\ell = \ep_\ell \ep_0 /t \)
of the pore-free membrane of thickness \(t\)\@.
There also is a small term due to the
surface tension \(\Sigma\) of the
plasma membrane of the cell, but
this term usually is negligible
since \(\Sigma\) is of the order of
\(2.5 \times 10^{-6}\) J/m\(^2\)~\citep{Dai1997}\@.
The energy of the pore in a plasma membrane is 
then~\citep{Chernomordik1979,Chernomordik1987,Chernomordik1988,Weaver1996,Chernomordik2001}
\beq
E(r) = 2 \pi r \gamma - \pi r^2 \Sigma 
- \thalf \pi r^2  \Delta C \, (\Delta V)^2.
\label {energy of pore}
\eeq
This energy has a maximum of
\beq
E(r_c) = \frac{\pi \gamma^2}{\Sigma + \thalf \Delta C (\Delta V)^2}
\approx \frac{2 \pi \gamma^2}{\Delta C (\Delta V)^2}
\label {Emax}
\eeq
at the critical radius 
\beq
r_c = \frac{\gamma}{\Sigma + \thalf \Delta C \, (\Delta V)^2}
\approx \frac{ 2 \gamma }{\Delta C \, (\Delta V)^2}.
\label {r critical}
\eeq
In Fig.~\ref{BolFac}, the Boltzmann
factor \(e^{-E(r)/(kT)}\) (\(\times 100\))
is plotted as a function of the radius \(r\)
of the pore up to \(r_c\)
for various transmembrane voltages
from \(- 200\) (solid, red) to \( -400\) mV
(dot-dash, cyan)\@.  Clearly,
the chance of a pore forming
rises steeply with the magnitude of the voltage
and falls with the radius of the pore.
\par
If the transmembrane potential \(\Delta V\) is turned off
before the radius of the pore reaches
\(r_c\), then the radius \(r\) 
of the pore usually shrinks quickly
(well within 1 ms~\citep{Chernomordik1987}) to a radius so small
as to virtually shut-down the conductivity
of the pore.  This rapid closure occurs
because in (\ref{energy of pore})
the energy \(2\pi r\gamma\) dominates
over \(- \pi r^2 \Sigma\), 
the surface tension
\(\Sigma\) being negligible.
Such a pore is said to be reversible.
But if \(\Delta V\) remains on when
\(r\) exceeds the critical
radius \(r_c\), then the pore usually
will grow and lyse the cell;
such a pore is said to be irreversible.
\par
\par
The formula (\ref{r critical}) provides
an upper limit on the radius of 
a reversible pore.  This upper limit
drops with the square of the transmembrane
voltage \(\Delta V\) from \(r_c = 3.6\) nm for \(\Delta V = - 200\) mV,
to 1.6 nm for \(\Delta V = - 300\), and 
to 0.9 nm for \(\Delta V = - 400\) mV\@.
\par
The time \( t_\ell \) for a pore's radius
to reach the critical radius \(r_c\) is 
the time to lysis; it
varies greatly and apparently randomly even
within cells of a given kind. 
In erythrocytes,
its mean value drops by nearly
an order of magnitude with each
increase of 100 mV in the
transmembrane potential~\citep{Chernomordik1987}
and is about a fifth of a second
when \(\Delta V = - 300\) mV\@.
\par
In the present model, however,
the potential is imposed by 
the CPP and the PSs, and so when
that potential causes a pore to form,
the CPP and its cargo may enter
the cell through the pore that they have
formed, and once they do, the potential
drops to its normal resting value,
usually less than -100 mV,
and the pore virtually closes
within 1 ms\@. 
\par
The oligoarginine(s) on the interface
between the outer leaflet and the
extra-cellular environment,
the negatively charged head groups 
of the PSs below them in inner leaflet,
and their counterions
create an electric field
and a transmembrane potential \(\Delta V\)\@.
The chance of this potential 
forming a pore of radius \(r\)
is proportional to
the Boltzmann factor \(e^{-E(r)/(kT)}\),
which is plotted in Fig.~\ref{BolFac}\@.
The higher the potential \(\Delta V\) and
the narrower the pore,
the greater the chance of pore formation.

\subsection{The Potential of an External Charge
\label{The Potential of an External Charge}}

\begin{figure}
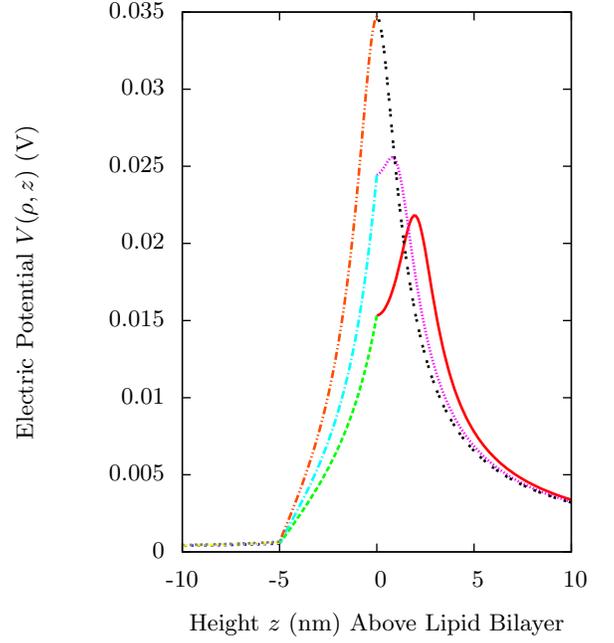

\centering
\input Vh
\caption{The electric potential \(V(\rho,z)\)
from (\ref{Vellt}--\ref{Vct})
in Volts for \(\rho = 1\) nm
as a function of the height \(z\) (nm) 
above the phospholipid bilayer 
(and in the extracellular environment)
for a unit charge \(q=|e|\) at 
\((\rho,z) = (0,0)\) (top curve), \((0,1)\)
(middle curve), and \((0,2)\) nm (bottom curve)\@.
The lipid bilayer extends from \(z=0\)
to \(z = -5\) nm, and the cytosol
lies below \(z = -5\) nm.
The relative permittivities 
were taken to be \(\ep_w = \ep_c = 80\)
and \(\ep_\ell = 2\)\@.}
\label {Vhfig}
\end{figure}
\begin{figure}
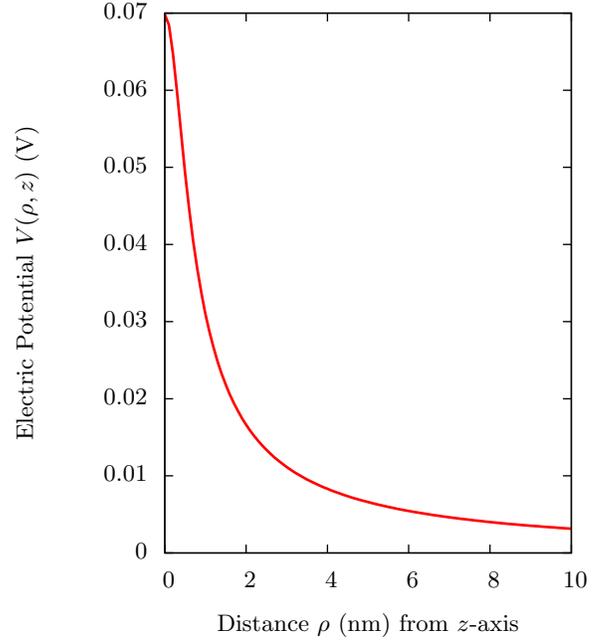

\centering
\input Vr
\caption{The electric potential \(V(\rho,z)\)
from (\ref{Vex}) in Volts 
for \(0 \le \rho \le 10\) nm
at a height \(z = 0.5\) (nm) 
above the phospholipid bilayer 
for a unit charge \(q=|e|\) at 
\((\rho,z) = (0,0)\) nm.
Same geometry and parameters as
in Fig.~\ref{Vhfig}\@.}
\label {Vrfig}
\end{figure}

As shown in Appendix~A,
the electrostatic potential 
in the lipid bilayer \(V_\ell(\rho,z)\)
due to a charge \(q\) at the point
\((0,0,h)\) on the \(z\)-axis a
height \( h \) above
the interface between the lipid bilayer 
and the extra-cellular environment is
\bea
V_\ell(\rho,z) & = &
\frac{q}{4\pi\epsilon_0 \epsilon_{w\ell}}
\, \sum_{n=0}^\infty
(p p')^n \lt(
\frac{1}{\sqrt{\rho^2 + (z-2nt-h)^2}} \rt. \nn\\
& & \lt. \mbox{} 
- \frac{p'}{\sqrt{\rho^2 + (z+2(n+1)t+h)^2}}\rt)
\label {Vellt}
\eea
for \(0 \le h \le t\),
in which \(t\) is the thickness
of the lipid bilayer,
\( \epsilon_{w\ell}
= ( \epsilon_w + \epsilon_\ell )/2 \)
is average of relative permittivity
of the extra-cellular fluid \( \epsilon_w \)
and that of the lipid bilayer \( \epsilon_\ell \),
and \(p\) and \(p'\) are the ratios
\beq
p = \frac{\epsilon_w - \epsilon_\ell}
{\epsilon_w + \epsilon_\ell}
\quad \mbox{and} \quad
p' = \frac{\epsilon_c - \epsilon_\ell}
{\epsilon_c + \epsilon_\ell}
\label {p and p' t}
\eeq
which lie between 0 and 1\@.
The potential in the extra-cellular medium is
\bea
V_w(\rho,z) & = & 
\frac{q}{4\pi\epsilon_0 \epsilon_w}
\lt(\frac{1}{r}
+ \frac{p}{\sqrt{\rho^2 + (z+h)^2}} 
\rt. \nn\\
& & \mbox{} -  \lt. 
\frac{\ep_w \ep_\ell}{\ep_{w\ell}^2}
\sum_{n=1}^\infty
\frac{p^{n-1} p^{\prime n}}
{\sqrt{\rho^2 + (z + 2nt + h)^2}} \rt)
\label {Vex}
\eea
in which \(r =\sqrt{\rho^2 + (z-h)^2}\) 
is the distance from the charge \(q\)\@.
The potential in the cytosol 
due to the same charge \(q\) is
\beq
V_c(\rho,z) = \frac{q \, \ep_\ell}
{4\pi\ep_0\ep_{w\ell}\ep_{\ell c}}
\! \sum_{n=0}^\infty
\frac{(p p')^n}{\sqrt{\rho^2 + (z-2nt-h)^2}}.
\label {Vct}
\eeq
where \(\ep_{\ell c}\) is
the mean relative permittivity
\(\ep_{\ell c} = ( \ep_\ell + \ep_c )/2\)\@.
\par
The first 1000 terms of the
series (\ref{Vellt}), (\ref{Vex}), \& (\ref{Vct})
for the potentials \(V_\ell(\rho,z)\),
\(V_w(\rho,z)\),
and \(V_c(\rho,z)\)
are plotted in Fig.~\ref{Vhfig} 
(in Volts) for \(\rho = 1\) nm
as a function of the height \(z\) (nm) 
above the phospholipid bilayer 
(and in the extracellular environment)
for a unit charge \(q=|e|\) at 
\((\rho,z) = (0,0)\) (top curve), \((0,1)\)
(middle curve), and \((0,2)\) nm (bottom curve)\@.
The lipid bilayer extends from \(z=0\)
to \(z = -5\) nm, and the cytosol
lies below \(z = -5\) nm.
The relative permittivities 
were taken to be \(\ep_w = \ep_c = 80\)
and \(\ep_\ell = 2\)\@.
Fig.~\ref{Vrfig} plots
the potential \(V_w(\rho,z)\)
in the extracellular region
due to a unit charge at the origin
as a function of \(\rho\)
for \(z = 0.5\) nm.
\par
In and near the extracellular region,
these potentials are fairly well
approximated by the simple formulas 
\bea
V_w(\rho,z) & \approx &
\frac{q}{4\pi\epsilon_0 \epsilon_w}
\lt(\frac{1}{r}
+ \frac{p}{\sqrt{\rho^2 + (z+h)^2}} \rt)
\label {Vw simple}\\
V_\ell(\rho,z) & \approx &
\frac{q}{4\pi \epsilon_0 \epsilon_{wl} \, r} 
\label {simple formula}
\eea
which hold when the lipid bilayer
is infinitely thick.  But the potential
drops significantly below this formula
(\ref{simple formula})
as \(z\) descends deeper into the
bilayer until it nearly vanishes
at the lipid-cytosol
interface and in the cytosol.
In fact, a charge of \(12|e|\) at
the origin raises
the potential \(V_\ell(\rho,z)\)
on the interface
at \((\rho,z) = (1,-5)\) nm
only to \(0.0079\) V\@.  Thus
the energy advantage of a PS at  
\((1,-5)\) nm
is only 0.0079 eV, which is much less
than \(kT_b \approx 0.027\) eV\@.
So a CPP
on the interface between the lipid
bilayer and the extracellular fluid
has a very small effect on the PSs
of the inner leaflet
whose negative charges lie on the
lipid-cytosol interface.  
\par
It follows that the counterions 
of the extracellular fluid also
have little effect upon the PSs.  
And since the electric
permittivities of the extracellular
fluid and of the cytosol are similar,
we may view Fig.~\ref{Vhfig} upside-down 
and conclude that the PSs and the K\(^+\)
and Cl\(^-\) ions of the
cytosol have little effect upon the
CPP and the counterions of the extracellular
fluid except to contribute most of the
transmembrane potential that exists
in the absence of CPPs.
Charges in the cytosol are
effectively decoupled from
those in the extracellular environment.
\par
We may draw a further lesson
from the sharp drop in \(V_\ell\)
across the lipid bilayer
shown in Fig.~\ref{Vhfig}:  The
transmembrane potential due
to a CPP on the interface (\(z=0\))
is much larger than the simple
formula (\ref{simple formula})
would imply.  This is why CPPs
are transduced.

\subsection{Monte Carlo of the Phosphatidylserines
\label{Monte Carlo of the Phosphatidylserines}}

\begin{figure}
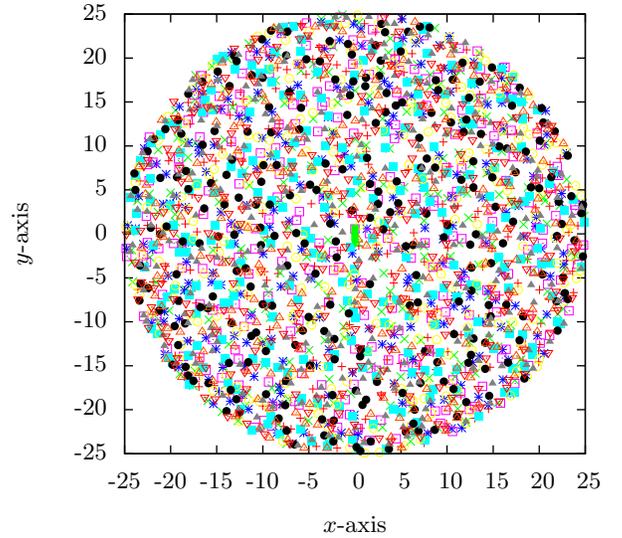

\centering
\input pspic
\caption{Superposition of 10 plots
of the positions \((x,y,-5)\) nm
of 255 phosphatidylserines inside a disk
of radius 25 nm on the lipid-cytosol
interface 5 nm below an \(\alpha\)-helix
of 12 arginines at \((0,y,0)\) 
for \(-0.96 \le y \le 0.8\) nm
(x's, green).  To a good approximation,
the PSs are uniformly and randomly
distributed.}
\label {PSfig}
\end{figure}

Phosphatidylserines (PSs) make up some 8--18\%
of the inner leaflet by weight~\citep{MBoC5.624}\@.
They diffuse laterally within
that leaflet with a diffusion
constant 
\(D \approx 10^{-8}\) 
cm\(^2\)/sec~\citep{MBoC5.622}
and so within one second 
spread to an area of 12 \(\mu\)m\(^2\),
which is a significant fraction
of the surface area of a eukaryotic cell.
\par
My Monte Carlo simulations of the
distribution of the PSs of the inner
leaflet verified the conclusions
of the last subsection 
(\ref{The Potential of an External Charge})
based upon the analytic potentials
(\ref{Vellt}--\ref{Vct})
and showed that the PSs are randomly
and uniformly distributed, at least
to a good approximation.  Figure~\ref{PSfig}
superposes 10 snapshots of the \((x,y)\) locations 
of 255 PSs  in a disk of radius 25 nm
that is 5 nm directly below a
12-mer of arginine R\(^{12}\)\@.
The snapshots were taken every 2000 sweeps
after 25,000 thermalizing sweeps.
The 255 PSs moved so as to minimize
their free energy due to interactions
with the R\(^{12}\), with each other,
and with the PSs outside the disk.
The distribution shows no obvious
clustering.
\par
I considered the case of
a single CPP 
of \(5 \le N \le 12\) arginines. 
A CPP of \( N \) arginines can form
an \( \alpha \)-helix of length
\( L_\alpha \approx 0.16 \, (N-1) \) nm,
a random coil of length
\( L_r \approx 0.25 \, (N-1) \) nm, 
or a \( \beta \)-strand of length 
\( L_\beta \approx 0.34 \, (N-1) \) nm.
The random-coil and \(\beta\)-strand
configurations spread the positively
charged guanidinium groups farther apart
and so would be expected to cluster
the PSs even less than the \(\alpha\)-helix
configuration.  So in the simulations
of this subsection, I only used
the \(\alpha\)-helix configuration.
\par
The Monte Carlo code~\citep{CahillCodes2009}
assumes that a PS has a cross-sectional
area of 1 nm\(^2\) and that the PSs make up
13\% of phospholipids of the inner leaflet.
There are then about \(0.13 \, \pi \, R^2\) PSs
in a disk of radius \(R\) nm, or
255 PSs in a disk of radius 25 nm.
The codes allow these 255 PSs to move
about within that disk attracted by
the electric potential of the
\(N\) or \(2 N\) positively charged arginines 
and repelled by each other and by
the PSs outside the disk, which are treated as
a uniform surface charge.  
The computations
are facilitated somewhat
by the continuity of the electric potentials 
(\ref{Vellt}, \ref{Vex}, \& \ref{Vct})
across the interfaces at \(z = 0\) and \(z = -t\)
between the lipid bilayer and respectively the
extra-cellular medium and the cytosol.
\par
The code assumes 
that the \(N\) arginines
form an alpha helix
with positive charges at the points
\beq
\bos{r}_{j CPP} = 
( 0, \,\,0.16 \, (N/2 - j), \,\, 0 ).
\label {r_cpp single CPP}
\eeq
The PSs were allowed to move
in two dimensions within the disk 
of radius 25 nm in the inner leaflet
at sites \(\bm{r}_k\) for \(k = 1, \dots 2N\)\@.
\par
The electrostatic energy of a single
PS at the point \(\bm{r}_k\) is the
sum of three different energies
\beq
E_k = E_{k,CPP} + E_{k,PSs} + E_{k,\sigma}.
\label {energy of PS}
\eeq
The first energy \(E_{k,CPP}\) is that due to its
interaction with arginines of the CPP(s)
\beq
E_{k,CPP} = \sum_{j=1}^M \mbox{} 
- e \, V_c(|\bm{r}_k - \bos{r}_{j CPP}|,-t)
\label {E PS CPP}
\eeq
in which \(M = N\) for a single CPP
and \(M = 2 N\) for two CPPs.
The second energy \(E_{k,PSs}\)
is that due to the interaction
of the \(k\)th \(PS\) with the \(N_{PS} -1 = 254\) 
other PSs in the disk
\beq
E_{k,PSs} =
\sum_{k'=1,k' \ne k}^{N_{PS}} e \, 
V_w(|\bm{r}_k - \bm{r}_{k'}|,0).
\label {E PS PSs}
\eeq
The third energy \(E_{k,\sigma}\) is
that due to the interaction
of the \(k\)th \(PS\) with all the
PSs outside the disk represented 
by a uniform surface charge 
\(\sigma = 0.13 \, e\) nm\(^{-2}\)\@.
In Appendix~B, I derive the approximation
\beq
E_{k,\sigma} \approx
\frac{\sigma R}{2 \ep_0 \ep_w} 
\sum_{n=1}^\infty \frac{1}{2n-1} 
\lt[ \frac{(2n)!}{(n!)^2 2^{2n}} \rt]^2  
\lt( \frac{r_k}{R} \rt)^{2n}
\label {E PS sigma}
\eeq
apart from an irrelevant infinite constant.
In the computer programs, the upper
limit on the summation was \(n = 800\)\@.
\par
The Monte Carlo codes use a simple
Metropolis step in which the \(x\)-\(y\) coordinates
of a single PS, the \(k\)th, are randomly varied
by as much as \(\pm 5 \) nm 
(to keep the acceptance rate down to 68\%)\@.
The codes accept any move
that lowers the energy \(E_k\) 
as given by (\ref{energy of PS})
and also accept any move
that raises \(E_k\) by \(\Delta E_k\) 
conditionally with probability
\beq
P = e^{- \Delta E_k /(k_B T)}
\label {P}
\eeq
in which \(k_B\) is Boltzmann's constant
and \(T\) is 37 Celsius.
A sweep consists of \(k = 1, \dots, N_{PS} = 255\)
Metropolis steps.
Each simulation started from a random
configuration of PSs in the disk of radius \(R = 25\) nm 
and ran for 45,000 sweeps.
Measurements began after
25,000 sweeps for thermalization.
\par
The radius of an electropore
is about \(r_p =1\) nm, and the thickness
of the plasma membrane was taken 
to be \(t = 5\) nm.
The code 
measured the electrostatic potential 
across the lipid bilayer between points
that were offset in the \(x\)-direction
by 1 nm, that is, between the points
\( ( 1, 0, 0 ) \) and \( ( 1, 0, -5 ) \) nm.
The code measured the voltage across
the membrane every 10 sweeps and 
recorded the positions of the PSs
every 2,000 sweeps.
\par
The transmembrane potential \(V\)
due to a single \(\alpha\)-helix
oligoarginine R\(^N\)
and its cloud of PSs
rises with the number of arginines
from about \(V = - 380\) mV  for \(N = 5\) arginines
to \(V = - 630\) mV  for \(N = 9\) 
and \(V = - 800\) mV  for \(N = 12\)\@.
These voltages are so high 
that one would have expected 
electroporation even for R\(^5\)s,
which is not seen.
The PSs contributed only about 70 mV
to the 
transmembrane potentials listed
in Table~\ref{tab:table of voltages}\@.
Thus, although
they facilitate transduction, 
they are not responsible for
the very high voltages listed
in the table.
These voltages are too high
because the simulations did
not include the Na\(^+\) and
Cl\(^-\) ions in the extra-cellular
medium.
\par
\par
In all these simulations, the mean value
of the distance of the PSs from the point
(0, 0, -5) nm was about 17 nm.  The distributions
of the PSs across the disk of radius \(r = 25\) nm
appeared uniform and random, with little clustering
under the CPPs as shown in Fig.~\ref{PSfig}\@.
The reason for the tepid PS response to
the electric field of the CPPs can be seen
in Fig~\ref{Vhfig}: the electric potential \(V(\rho,z)\)
drops off sharply as \(z\) descends through the lipid
bilayer and is very small near the lipid-cytosol interface.
This uniformity of the PS distribution
on the inner leaflet means that we need
not simulate their behavior explicitly.
We can use the resting transmembrane potential
in the absence of CPPs 
to represent both the PSs and the
counterions of the cytosol.
Ideally, one should take it
from experimental measurements.

\subsection{The Potential of an Oligoarginine
\label{The Potential of an Oligoarginine}}

This subsection computes
the transmembrane potential \(\Delta V_{CPP}\)
due to an R\(^n\) oligoarginine 
whose \(n\) unit positive charges for
\(5 \le n \le 12\) were fixed at
the points 
\beq
\bos{r}_{j CPP} = 
( 0, \,\, (N/2 - j) \, \Delta y, \,\, 0 )
\label {r_cpp single bCPP}
\eeq
in which \(\Delta y = 0.16\), 0.25,
and 0.34 nm respectively
for an \(\alpha\)-helix,
a random coil,
and a \(\beta\)-strand.
I took this \(\Delta V_{CPP}\)
to be the difference
\beq
\Delta V_{CPP} = \la V_c(\rho,-t) \ra
- \la V_w(\rho,0) \ra
\label {Vcpp}
\eeq
in which \(\la V_w(\rho,0) \ra\)
and \(\la V_c(\rho,-t) \ra\) are
the mean values of the  R\(^n\)'s 
electric potential
on two disks of radius \(r_p = 1\) nm
at \(z=0\) and at \(z = -t\)\@.
\par
I used a Monte Carlo code~\citep{CahillCodes2009}
to numerically integrate the
appropriate potential
\(V_w\) or \(V_c\) (Eqs.~(\ref{Vex} or \ref{Vct}))
over the two disks.
The code used a million 
random points on each of the disks
(of which the fraction \((4-\pi)/4 = 0.215\)
were discarded because they lay outside
the disk).
In this code,
I kept 100 terms in the series
(\ref{Vex} \& \ref{Vct});
the error introduced by
this truncation is completely
negligible (about
2 parts in 10 million)\@.
\par
The resulting transmembrane potentials
\(\Delta V_{CPP}\) are listed in 
Table~\ref{tab:R^n table of voltages}\@.
The magnitude of \(\Delta V_{CPP}\)
naturally increases with the charge \(n|e|\)\@.
Because the \(n\) charges
are more spread out in a \(\beta\)-strand
than in a random coil, the magnitude of \(\Delta V_{CPP}\)
is less for a \(\beta\)-strand than for a random coil
of the same charge, and similarly
for a coil and an \(\alpha\)-helix.

\par
\begin{table}
\caption{The voltage differences 
\( \Delta V_{CPP} \) (mV)
across the plasma membrane due to an R\(^N\)
oligoarginine as an \(\alpha\)-helix,
a random coil, or a \(\beta\)-strand.  
Neither the salt potential \(\Delta V_{NaCl}\)
nor the resting cell potential
\(\Delta V_{cell}\) is included.}
\label {tab:R^n table of voltages}
\begin{ruledtabular}
\begin{tabular}{||c|c|c|c||} 
\, \(N\) 
& R\(^N\) \(\alpha\)-helix 
& R\(^N\) random coil
& R\(^N\) \(\beta\)-strand 
\\ \hline 
5 &  \(\mbox{}-312 \)   & \(\mbox{}-302\)   & \(\mbox{}-291\) \\ \hline
6 &  \(\mbox{}-376 \)   & \(\mbox{}-362\)  & \(\mbox{}-346\) \\ \hline
7 &  \(\mbox{}-439 \)  & \(\mbox{}-419\)    & \(\mbox{}-393\) \\ \hline
8 &  \(\mbox{}-502 \)  & \(\mbox{}-472\)    & \(\mbox{}-425\) \\ \hline
9 &  \(\mbox{}-562 \) & \(\mbox{}-521\)    & \(\mbox{}-455\) \\ \hline
10 & \(\mbox{}-620 \)  & \(\mbox{}-557\)   & \(\mbox{}-476\) \\ \hline
11 & \(\mbox{}-676 \)   & \(\mbox{}-587\)   & \(\mbox{}-499\) \\ \hline
12 & \(\mbox{}-729 \) & \(\mbox{}-614\)    & \(\mbox{}-516\) \\ \hline
\end{tabular}
\end{ruledtabular}
\end{table}

\subsection{Monte Carlo of the Counterions
\label{Monte Carlo of the Counterions}}

In this subsection, I use Monte Carlo
methods to compute the transmembrane 
potential \(\Delta V_{NaCl}\) due to
the sodium and chloride ions of 
the extracellular medium near
an oligoarginine.
\par 
The Na\(^+\), K\(^+\), Mg\(^{++}\),
Ca\(^{++}\), and Cl\(^-\) concentrations
in the extracellular medium
respectively are 145, 5, 1--2, 1--2,
and 110 mM~\citep{MBoC5p652}\@.
I approximated their effects by setting
the Na\(^+\) and Cl\(^-\) 
concentrations to 156 mM
and ignoring the other ions.
I used an active volume
that was 10 nm wide and 20 nm long,
and that rose from
the lipid bilayer to a height of 5 nm.  
In this active volume
of 200 (nm)\(^3\), I put
94 sodium ions and (94+n)
chloride ions so as to make the
charge within the active volume neutral.
\par
To prevent the sodium and chloride
ions from avoiding the walls and
ceiling of the active volume,
I surrounded the walls and
ceiling of the active volume
with a 1000 (nm)\(^3\) 5 nm-thick
passive volume in which I randomly
placed 470 Na\(^+\) and 470 Cl\(^-\) ions.
\par
The Monte Carlo code~\citep{CahillCodes2009}
used the potential 
\(V_w\) of Eq.~(\ref{Vex})
to compute the energy of an individual
sodium or chloride ion in the active
volume due to its interaction
with all the ions in the active
and passive volumes and with the
CPP(s) which did not move.  
The fixed positions
(\ref{r_cpp single bCPP})
of the \(n\) charges 
of the oligoarginine depended upon whether the R\(^n\)
was configured as an \(\alpha\)-helix,
a random coil,
or a \(\beta\)-strand.
The ions in the passive volume also
didn't move,
retaining their original random positions,
which were different in each run.
To speed up the computation, I used
only the first 8 terms in the series
(\ref{Vex}) for \(V_w(\rho,z)\),
which introduced an error of about 0.6\%\@.
\par
In order to prevent the Na\(^+\) and
Cl\(^-\) ions from collapsing into
neutral composite particles of infinite
negative energy, I added to \(V_w(\rho,z)\)
the hard core
\beq
V_{NaCl}(r) = \frac{e}{4\pi\epsilon_0 \epsilon_w} \,
\frac{r_0^{11}}{12 r^{12}}.
\label {VNaCl}
\eeq
If we keep only the \(1/r\) term
of \(V_w(\rho,z)\), then 
the potential \(V_w + V_{NaCl}\) is 
proportional to
\beq
\frac{r_0^{11}}{12 r^{12}} - \frac{1}{r}
\label {VNaCl2}
\eeq
which has a minimum at \(r = r_0\)\@.
I took this parameter to be \(r_0 = 0.51\) nm 
which is the location of both the outer
maximum of the NaCl-in-water correlation function
\(g(r)\) and also the outer 
minimum of the (\tsc{scpism} plus \tsc{sif}) 
potential energy
of Na\(^+\)--Cl\(^-\) in water~\citep{Hassan2008}.
This choice of \(r_0\) allows the 
Na\(^+\) and Cl\(^-\) ions
to keep their hydration shells;
97\% of them do keep their
hydration shells at 100 mM 
and 25 C\@~\citep{Hassan2008}.
To prevent the chloride ions from
falling into the positive charges of
the arginines, I added a similar term
to the R--Cl potential but used the
somewhat larger value of 
\(r_0 = 0.7\) nm to account 
for the more spread-out charge
of the bidentate guanidinium group.
\par
Since I treated the water and the
lipids as dielectrics, I took the
potential energy of a sodium or
chloride ion of charge \(q=\pm |e|\)
to be proportional to the transmembrane
voltage \(V_{cell}\) reduced by the
ratio of the two permittivities
and by the ratio of the height \(z\)
to the thickness \(t\) of the lipid
bilayer
\beq
V(z) = q \, \frac{\ep_\ell}{\ep_w} \, 
\frac{z}{t} \, V_{cell}.
\label {V due to Vcell}
\eeq
This energy is small compared to \(kT_b\)\@.
Even for \(V_{cell} = 100\) mV and \(z = t\),
it's only \(kT_b/10\)\@.  I used
the nominal value of 60 mV for \(V_{cell}\)
in my simulation of the effect of the
salt on the transmembrane potential.
\par
The Monte Carlo code measured 
the transmembrane potential 
\beq
\Delta V_{NaCl} = V_{NaCl}(0,0,-t)
- V_{NaCl}(0,0,0)
\label {dVsalt}
\eeq
due to the salt ions of the active volume.
It used the first 100 terms
of the potential \(V_c(\rho,z)\)  
in the cytosol (\ref{Vct}) to
compute \( V_{NaCl}(0,0,-t) \),
and it used the first 100 terms
of the potential \(V_w(\rho,z)\)
in the extracellular medium (\ref{Vex})
to compute \( V_{NaCl}(0,0,0) \)\@.
The errors of truncation were negligible.
\par
Each run started by assigning
random positions to the 94
Na\(^+\) ions and the \((94 + n)\)
Cl\(^-\) ions of the active volume
and to the 470 sodium and 470 chloride
ions of the passive volume.
After this initialization,
the code did 25,000
thermalizing sweeps in which
every Na\(^+\) and Cl\(^-\) ion
of the active volume was allowed
to move as much as \(1/4\)th
of its range in each direction.
After thermalization, the code
measured the transmembrane
potential every 10 sweeps
for a total of 2500 measurements.
Five runs were done for each number
\(5 \le n \le 12\) of arginines.
The resulting transmembrane potentials
\(\Delta V_{NaCl}\) due to the salt
are listed in mV in 
Table~\ref{tab:NaCl}\@.
\par
\begin{figure}
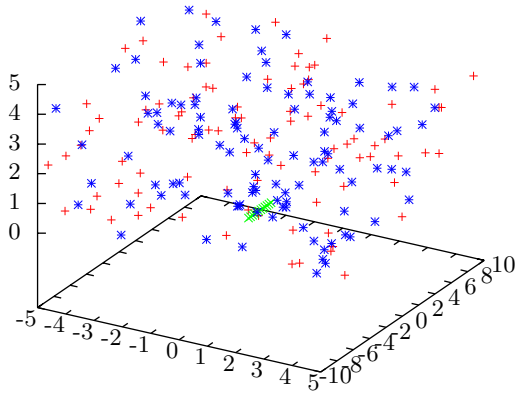

\centering
\input NaCl
\caption{Snapshot of 94 sodium ions (pluses, red),
106 chloride ions (stars, blue),
and a 12-mer random coil of oligoarginine
R\(^{12}\) (x's, green) after 50,000 sweeps.
The coordinates are in nm.}
\label {NaClfig}
\end{figure}

The code took snapshots
of the distributions of
the sodium and chloride ions
every 2500 sweeps after thermalization.
Fig.~\ref{NaClfig} displays the last
snapshot (after 50,000 sweeps) of 94 Na\(^+\),
106 Cl\(^-\), and 12 Rs in a random coil.
The coordinates are in nm.

\par
\begin{table}
\caption{The voltage differences 
\( \Delta V_{NaCl} \) (mV)
due to the 156 mM Na\(^+\) and Cl\(^-\) ions
near an R\(^N\)
oligoarginine as an \(\alpha\)-helix,
a random coil, 
or a \(\beta\)-strand. 
The resting transmembrane potential
\(\Delta V_{cell}\) is not included,
nor that \(\Delta V_{CPP}\) due to 
the oligoarginine.}
\label {tab:NaCl}
\begin{ruledtabular}
\begin{tabular}{||c|c|c|c||} 
\, \(N\) 
& R\(^N\) \(\alpha\)-helix 
& R\(^N\) random coil
& R\(^N\) \(\beta\)-strand 
\\ \hline 
5 &  \(\mbox{}168 \pm 4\) & \(\mbox{}148 \pm 3\) & \(\mbox{}143 \pm 4\) \\ \hline
6 &  \(\mbox{}202 \pm 4\) & \(\mbox{}189 \pm 4\) & \(\mbox{}178 \pm 1\) \\ \hline
7 &  \(\mbox{}233 \pm 4\) & \(\mbox{}218 \pm 3\) & \(\mbox{}204 \pm 2\) \\ \hline
8 &  \(\mbox{}270 \pm 3\) & \(\mbox{}244 \pm 1\) & \(\mbox{}226 \pm 5\) \\ \hline
9 &  \(\mbox{}306 \pm 6\) & \(\mbox{}275 \pm 2\) & \(\mbox{}250 \pm 4\) \\ \hline
10 & \(\mbox{}339 \pm 2\) & \(\mbox{}297 \pm 5\) & \(\mbox{}266 \pm 4\) \\ \hline
11 & \(\mbox{}370 \pm 5\) & \(\mbox{}327 \pm 3\) & \(\mbox{}281 \pm 4\) \\ \hline
12 & \(\mbox{}406 \pm 4\) & \(\mbox{}353 \pm 4\) & \(\mbox{}303 \pm 2\) \\ \hline
\end{tabular}
\end{ruledtabular}
\end{table}
\par
\begin{table}
\caption{\label{tab:1-cpp table of voltages}
The voltage differences 
\( \Delta V_{CPP} + \Delta V_{NaCl} \) (mV)
across the plasma membrane induced by an R\(^N\)
oligoarginine as an \(\alpha\)-helix,
a random coil, 
or a \(\beta\)-strand and by the ions
of 156 mM Na\(^+\) and Cl\(^-\)
reacting to it.
The resting transmembrane potential
\(\Delta V_{cell}\) is not included.}
\label {1-cpp table of voltages with counterions}
\begin{ruledtabular}
\begin{tabular}{||c|c|c|c||} 
\, \(N\) 
& R\(^N\) \(\alpha\)-helix 
& R\(^N\) random coil
& R\(^N\) \(\beta\)-strand 
\\ \hline 
5 &  \(\mbox{}-144 \pm 4\)     & \(\mbox{}-154 \pm 3\)   & \(\mbox{}-148 \pm 4\) \\ \hline
6 &  \(\mbox{}-174 \pm 4\)   & \(\mbox{}-173 \pm 4\)  & \(\mbox{}-168 \pm 1\) \\ \hline
7 &  \(\mbox{}-206 \pm 4\)  & \(\mbox{}-201 \pm 3\)    & \(\mbox{}-189 \pm 2\) \\ \hline
8 &  \(\mbox{}-232 \pm 3\)  & \(\mbox{}-228 \pm 1\)    & \(\mbox{}-199 \pm 5\) \\ \hline
9 &  \(\mbox{}-256 \pm 6\) & \(\mbox{}-246 \pm 2\)    & \(\mbox{}-205 \pm 4\) \\ \hline
10 & \(\mbox{}-281 \pm 2\)  & \(\mbox{}-260 \pm 5\)   & \(\mbox{}-210 \pm 4\) \\ \hline
11 & \(\mbox{}-306 \pm 5\)   & \(\mbox{}-260 \pm 3\)   & \(\mbox{}-218 \pm 4\) \\ \hline
12 & \(\mbox{}-323 \pm 4\) & \(\mbox{}-261 \pm 4\)    & \(\mbox{}-213 \pm 2\) \\ \hline
\end{tabular}
\end{ruledtabular}
\end{table}
\par

\subsection{Summary of the Model
\label{Summary of the Model}}

In the present model of CPP transduction,
the transmembrane potential is the sum
of three terms---the resting transmembrane
potential \(\Delta V_{cell}\) of the cell
in the absence of CPPs, the transmembrane
potential \(\Delta V_{CPP}\) due to an
oligoarginine, and the  transmembrane
potential \(\Delta V_{NaCl}\) due to
the counterions of the extracellular medium
\beq
\Delta V = \Delta V_{cell} + 
\Delta V_{CPP} + \Delta V_{NaCl}.
\label {dV3}
\eeq
The dominant term is the one \(\Delta V_{CPP}\)
due to the CPP; it is nearly twice as big 
as the one \(\Delta V_{NaCl}\) due to the
salt and of opposite sign.  
The resting transmembrane
potential \(\Delta V_{cell}\) of the cell,
which arises mostly from the phosphatidylserines
of the inner leaflet, augments the sum
\(\Delta V_{CPP} + \Delta V_{NaCl}\)
by some 10--50\% depending upon the CPP's
charge and the value of \(\Delta V_{cell}\)\@.
This salty CPP-PS capacitor increases the
transmembrane potential \(V\)
and so elevates the
Boltzmann factor \(e^{-E(r)/(kT)}\) and so
increases the probability of
pore formation---at least for
R\(^N\)s with enough arginines.
It is hard to be quantitative here
because the voltage required to form
a pore depends upon the duration
of the voltage, the radius of the pore,
and any defects or fluctuations in the membrane.
\par
In its use of an electric field
and of the binding of the CPPs
to the phosphate groups of the phospholipids
of the outer leaflet,
the model has something in common with
the adaptive-translocation model 
of Rothbard, Jessop, and Wender~\citep{Wender2005};
in its invocation of electroporation,
it has some overlap with the work
of Binder and Lindblom~\citep{Lindblom2003};
in its use of 
neutral dipolar PC \& SM head groups
it is somewhat similar to the work of 
Herce and Garcia~\citep{Garcia2007}
and of Tang, Waring, and Hong~\citep{Hong2007}\@.
The key distinctive feature
of the present model is its
underpinning of continuum
electrostatics and its quantitative synthesis
of the contributions of the CPP,
the salt, and the phosphatidylserines
which combine to form 
a salty CPP-PS capacitor with a voltage
high enough to cause reversible 
electroporation.

\section{Comparison with Experiment
\label{Comparison with Experiment}}

\subsection{Empirical Upper Limit on Size of Cargo
\label{Empirical Upper Limit on Size of Cargo}}

Various groups have found that
cell-penetrating peptides cannot
transduce cargos of more than 
about 50 amino acids~\citep{Cardoso2006},
an upper limit that surely varies with
the cell, the CPP, and the cargo\@.  
In the transduction experiments~\citep{Tsien2004,Dowdy2005,Pugh2002,Kaelin1999,Fong2003,Cohen2002,Robbins2001,Datta2001,Hosotani2002,Hsieh2006,SnyderDowdy2004,Morano2006,Tuennemann2007} aimed
at eventual therapies,
the heaviest cargo was 33 amino acids.
The present model based on molecular
electroporation offers a qualitative
explanation for this upper limit.
\par
The masses of 
larger cargoes
of \(A = 50\)--100 amino acids
of 130 Da each together
with \(N\) arginines
have masses \(M_{N,A}\)
of from \(M_{N,A} = 0.1562 N + 6.5 \)
to \(M_{N,A} = 0.1562 N + 13 \) kDa.
Our previous formula (\ref{r =})
gives the lower bounds on the radii
of such proteins that run from
1.29 to 1.58 nm for CPPs of \(N = 10\) 
arginines.
But the energy \(E(r)\) of a pore
rises with its radius \(r\) as shown
by Eq.(\ref{energy of pore}) and so
the chance of pore formation falls
with the pore radius as shown by
Fig.~\ref{BolFac}\@.
So the chance of a pore forming
that is big enough for a cargo
much larger than 50 aa is small.
Such cargoes can't easily
fit through the pores that
are most likely to form.

\subsection{Experiments with Mouse Myoblasts
\label{Experiments with Mouse Myoblasts}}

\par
\begin{table}
\caption{\label{tab:table of mouse data}
The fractions of mouse C\(_2\)C\(_{12}\)
myoblasts transduced by oligoarginines
L-R\(^N\) at three concentrations (\(\mu\)M).}
\label {table of mice}
\begin{ruledtabular}
\begin{tabular}{||c|c|c|c||} 
\, \(N\) 
& 10 \(\mu\)M & 50 \(\mu\)M & 100 \(\mu\)M \\ \hline 
5 &  \(0.0 \pm 0.01\)     & \(0.0 \pm   0.01\)  & \(0.01 \pm   0.01\) \\ \hline
6 &  \(0.0 \pm 0.01\)     & \(0.04 \pm  0.03\)  & \(0.28\pm   0.03\) \\ \hline
7 &  \(0.0 \pm 0.01\)     & \(0.18\pm   0.03\)  & \(0.75 \pm  0.03
\) \\ \hline
8 &  \(0.02 \pm 0.02\)     & \(0.31 \pm  0.06\)  & \(0.85 \pm  0.04\) \\ \hline
9 &  \(0.05 \pm 0.04\)     & \(0.42\pm   0.10\)  & \(0.90 \pm  0.04\) \\ \hline
10 & \(0.70 \pm 0.04\)     & \(0.73\pm   0.04\)  & \(0.91 \pm  0.04\) \\ \hline
11 & \(0.81 \pm 0.07\)     & \(0.83 \pm  0.03\)  & \(0.92\pm   0.04\) \\ \hline
12 & \(0.90 \pm 0.04\)     & \(0.92 \pm  0.02\)  & \(0.99 \pm  0.02\) \\ \hline
\end{tabular}
\end{ruledtabular}
\end{table}

\begin{table}
\caption{\label{tab:table of voltages}
The voltage differences \( \Delta V \) (mV)
across the plasma membrane induced by two
oligoarginines separated by 2 nm
and configured as \(\alpha\)-helices
or as \(\beta\)-strands.  The solution
was 156 mM NaCl as in 
Table~\ref{1-cpp table of voltages with counterions}\@.
Includes a resting potential \(\Delta V_{cell} = - 30\) mV\@.}
\begin{ruledtabular}
\begin{tabular}{||c|c|c||} 
\, \(N\) 
& 2 R\(^N\) \(\alpha\)-helices
& 2 R\(^N\) \(\beta\)-strands  \\ \hline 
5 &  \(\mbox{}-249 \pm 3\)     & \(\mbox{}-238 \pm 4\)  \\ \hline
6 &  \(\mbox{}-286 \pm 7\)     & \(\mbox{}-271 \pm 6\)  \\ \hline
7 &  \(\mbox{}-333 \pm 5\)     & \(\mbox{}-284 \pm 4\)  \\ \hline
8 &  \(\mbox{}-365 \pm 6\)     & \(\mbox{}-295 \pm 3\)  \\ \hline
9 &  \(\mbox{}-401 \pm 11\)     & \(\mbox{}-304 \pm 4\) \\ \hline
10 & \(\mbox{}-447 \pm 3\)     & \(\mbox{}-308 \pm 7\)  \\ \hline
11 & \(\mbox{}-483 \pm 2\)     & \(\mbox{}-310 \pm 6\)  \\ \hline
12 & \(\mbox{}-509 \pm 5\)     & \(\mbox{}-308 \pm 5\)  \\ \hline
\end{tabular}
\end{ruledtabular}
\end{table}

T\"unnemann \textit{et al.}~\citep{Cardoso2008}
used  confocal
laser scanning microscopy to measure
the ability of
the L- and D-isoforms of oligolysine and of
oligoarginine to carry fluorophores 
of \(\sim\!\!400\) Da into live C\(_2\)C\(_{12}\)
mouse myoblasts within one hour. 
They found that oligoarginines transduced 
the fluorophores much better than oligolysines
and that more arginines meant faster transduction,
with L-R9 and L-R10 doing better than 
their shorter counterparts as shown
in Table~\ref{table of mice}\@.
They also found that
the D-isoforms worked better than the L-isoforms
and that transduction
rose with the CPP concentration
faster than linearly, which may
suggest a cooperative effect.
\par
The present model is consistent
with these experimental facts and
explains them as follows:
The oligoarginines crossed cell membranes
more easily than the oligolysines
because they were better able to bind to
the phosphate groups of the PCs and
SMs in the outer leaflet; the
oligolysines were not able to
form a stable upper plate of
a salty CPP-PS capacitor.
CPPs with more arginines were
transduced more rapidly because
with more arginines they could
bind to more PCs and SMs and because
their higher charges led to higher
transmembrane potentials, as noted
in Table~\ref{tab:R^n table of voltages}\@. 
The D-isoforms worked better than the L-isoforms
because the capacitor mechanism
is insensitive to the chirality
of the amino acids and because
proteases were less able to cut them.
To check for a cooperative effect,
I ran some Monte Carlo simulations in which
two oligoarginines were as close as 2nm.
In these simulations, I set \(r_0 = 0.55\) nm
for NaCl and 0.7 nm for R-Gdm.
The resulting transmembrane potentials \(\Delta V\)
are listed in Table~\ref{tab:table of voltages}
for \(\Delta V_{cell} = -30 \) mV\@.
They are higher than those due to a single R\(^n\),
which appear in 
Table~\ref{1-cpp table of voltages with counterions}
(even after \(\Delta V_{cell} = -30 \) mV is added in)\@.
Thus higher CPP concentrations accelerate
transduction because they increase the odds
of two or more CPPs attaching to nearly
the same spot on the outer leaflet.
There is also the \tit{possibility}
that under physiological conditions
two oligoarginines might form an
anti-parallel \(\beta\)-sheet~\citep{Jungwirth2009}\@.
Such \(\beta\)-sheets would entail
a cooperative effect.
\par
This consistency of the capacitor model
and its simplicity
lends it some plausibility.  
But evolution finds what works, 
not what fits neatly into a model, 
and so other CPPs with
different cargos may enter different
cells by different mechanisms.
In particular, this model may not
apply to model amphiphilic peptides (MAPs)\@.

\section{Three Tests of the Model
\label{Three Tests of the Model}}

One way to test the model
would be to compare the 
rates of polyarginine 
transduction 
in wild-type cells and in those that
have little or no phosphatidylserine (PS) 
in their plasma membranes.
If PS plays a role as in the model
of this paper and augments the
transmembrane potential by 10--50\%,
then the transduction of
polyarginine fused to a cargo
of less than 30 amino acids
should be somewhat faster in the wild-type cells
than in those without PS in their
plasma membranes.
Mammalian cell lines that are
deficient in the synthesis
of phosphatidylserine do 
exist~\citep{Frazier1986,Nishijima1996,Kuge1998,Vance2000,Vance2004},
but they appear to have normal levels
of PS in their 
plasma membranes~\citep{Vance2004}---presumably due
to a lower rate of PS degradation~\citep{Vance2008}\@.
\par
Another test 
would be to construct artificial asymmetric 
bilayers~\citep{Schlue1993,Bezrukov1998,Ligler2007} 
with and without PS
on the ``cytosolic'' side and to compare
the rates of CPP-cargo transduction.
If the present model is right,
then the rate of transduction
should be somewhat higher through membranes
with PS on the cytosolic side than
through membranes with no PS or with
PS on both sides.
\par
If CPPs do enter cells via molecular
electroporation, then 
it may be possible 
to observe the formation
of transient (\(< 1\) ms) pores by 
detecting changes in the conductance
of the membrane~\citep{Chernomordik2001}\@.
Such measurements would be
a key test of the model 
and, if done on an artificial membrane,
would let one 
determine both
whether CPP-transduction
is related to the presence
of PS on the cytosolic side of the
membrane and whether it
proceeds via molecular
electroporation.

\vspace{.2in}

\section{Summary
\label{Summary}}

Cell-penetrating peptides (CPPs) can carry into cells
cargoes with molecular weights 
of as much as 3,000 Da---much greater than 
the nominal limit 500 of the 
``rule of 5''~\citep{Lipinski1997}\@.
Therapeutic applications with well-chosen
peptide cargoes of 8--33 amino acids
are described in references~\citep{Tsien2004,Dowdy2005,Pugh2002,Kaelin1999,Fong2003,Cohen2002,Robbins2001,Datta2001,Hosotani2002,Hsieh2006,SnyderDowdy2004,Morano2006,Tuennemann2007}\@.
\par
Sec.~\ref{The Model}
describes a model in which 
molecular electroporation and phosphatidylserines (PSs)
play key roles
in the transduction of CPP-cargo molecules.
In this model, one or more positively charged CPPs
on the outer leaflet and the negatively charged
PSs under it on the inner leaflet form a kind 
of capacitor with a transmembrane potential
in excess of 180 mV
for a single CPP of nine arginines.
This transmembrane potential increases the
chance of the formation of electropores
through which the CPP and its cargo
can enter the cell.
The model is consistent with
the empirical upper limit 
on the cargo of about
50 amino acids and with
data~\citep{Cardoso2008}
on how the probability of transduction
of polyarginine CPPs into mouse myoblasts
depends upon the 
concentration of the CPP-cargo
molecules and the number of 
arginines in each CPP\@.
\par
The model predicts that
mammalian cells that lack
phosphatidylserine in their
plasma membranes
transduce polycations less well
than those that do, 
that artificial asymmetric 
bilayers with PS
on the cytosolic side 
transduce polycations better than
ones without PS, and that
the passage of CPPs should be
accompanied by transient rises
in the conductance of the membrane
of the cell or BLM\@. 

\appendix
\section{First Electrostatic Problem
\label{First Electrostatic Problem}}

Here we derive in the continuum limit
the electrostatic potential \(V(\rho,\phi,z)\) in
cylindrical coordinates 
due to a charge \(q\) on the \(z\)-axis
at the point \((0,0,h)\) at a height
\(h\) in the extracellular environment
above the phospholipid bilayer of a eukaryotic cell
for the case in which the height \(h\)
does not exceed the thickness \(t\)
of the lipid bilayer.
\par
In electrostatic problems,
Maxwell's equations reduce to Gauss's law
\beq
\bos{\nabla \cdot \mbf{D}} = \rho
\label {VD=r}
\eeq
which relates the divergence
of the electric displacement \(\mbf{D}\)
to the  density \( \rho \)
of free charges
(charges that are free to move in or out of
the dielectric medium---as opposed to those 
that are part of the medium and bound
to it by molecular forces),
and the static form of Faraday's law
\beq
\bos{\nabla \times \tbf{E}} = 0
\label {vxE=0}
\eeq
which implies that the electric
field \( \mbf{E} \) is the gradient 
of an electrostatic potential
\beq
\mbf{E} = - \bos{\nabla} V.
\label {E=-VV}
\eeq 
\par
Across an interface with normal vector \( \mbf{\hat n} \)
between two dielectrics,
the tangential component of the
electric field is continuous
\beq
\mbf{\hat n} \times 
\lt( \mbf{E}_2 - \mbf{E}_1 \rt) = 0
\label {nx(E2-E1)=0}
\eeq
while the normal component
of the electric displacement
jumps by the surface density \( \sigma \)
of free charge
\beq
\mbf{\hat n} \cdot
\lt( \mbf{D}_2 - \mbf{D}_1 \rt) = \sigma.
\label {dD = sigma}
\eeq
\par
In a linear dielectric,
the electric displacement \(\bos{D}\)
is proportional to the electric field \(\bos{E}\)
\beq
\bos{D} = \epsilon \, \bos{E}
\label {D=eE}
\eeq
and the coefficient \( \epsilon \) 
is the permittivity
of the material.  The permittivity
\( \epsilon(m) \) of a material
\( m \) differs from
that of the vacuum \( \epsilon_0 \) by the
electric susceptibility \( \chi \)
and by the relative permittivity
\( \epsilon_m \)
\beq
\epsilon(m) = \epsilon_0 + \chi 
=  \epsilon_m \, \epsilon_0 .
\label {eps = eps0 + chi}
\eeq
The relative permittivity \( \ep_m \)
often is denoted \(K_m\)\@.
\par
The lipid bilayer
is taken to be flat 
and of a thickness
\(t \approx 5\)~nm.  The relative permittivity
of the lipid bilayer is 
\(\epsilon_\ell \approx 2\),
that of the extra-cellular environment
is \(\epsilon_w \approx 80\), and that
of the cytosol is \(\epsilon_c \approx 80\)\@.
\par
We use the method of image charges.
The charge \(q\) at \((0,0,h)\)
will generate image charges at the
points \(\bos{r} = (0,0,2nt\pm h)\)
in which \(n\) runs
over all the integers.  
The cylindrical symmetry of the problem
ensures that the potential is independent
of the azimuthal angle \(\phi\) and 
so can depend only upon \(\rho\) and \(z\)\@.
With \(\rho^2 = x^2 + y^2\), the
potential in the lipid bilayer is
\bea
V_\ell(\rho,z) & = & 
\frac{1}{4\pi\epsilon_0 \epsilon_\ell}
\lt[ 
\frac{q_{0}}{\sqrt{\rho^2 + (z - h)^2}}
\rt. \label {Vlipidh}\\
& & \lt. + \sum_{s=\pm1} 
\sum_{0 \ne n=-\infty}^\infty
\frac{q_{n,s}}
{\sqrt{\rho^2 + (z - (2nt+sh))^2}}\rt]
\nn
\eea
while that in the extracellular
environment is
\beq
V_w(\rho,z) = 
\frac{1}{4\pi\epsilon_0 \epsilon_w}
\sum_{s=\pm1} \sum_{n=-\infty}^0
\frac{q_{wn,s}}
{\sqrt{\rho^2 + (z - (2nt+sh))^2}}
\label {Vexth}
\eeq
and that in the cytosol is
\beq
V_c(\rho,z) = 
\frac{1}{4\pi\epsilon_0 \epsilon_c}
\sum_{s=\pm1} \sum_{n=0}^\infty
\frac{q_{cn,s}}
{\sqrt{\rho^2 + (z - (2nt+sh))^2}}.
\label {Vcyth}
\eeq
\par
The continuity (\ref{nx(E2-E1)=0}) 
of the transverse electric 
field \(E_\rho\) and that (\ref{dD = sigma})
of the normal displacement \(D_z\) 
across the planes \(z=0\) and \(z = -t\)
imply that the coefficients
\(q_0\), \(q_{n,s}\), \(q_{wn,s}\), and
\(q_{cn,s}\) must satisfy 
for \(n>0\) and \(s=\pm1\)
the relations
\bea
q_{n,s} + q_{-n,-s} & = & \frac{\ep_\ell}{\ep_w}
q_{w-n,-s} \label {410}\\
q_{n,s} - q_{-n,-s} & = & - q_{w-n,-s} \\
q_{n,s} + q_{-(n+1),-s} & = & \frac{\ep_\ell}{\ep_c}
q_{cn,s} \\
q_{n,s} - q_{-(n+1),-s} & = & q_{cn,s} 
\label {440}
\eea
as well as the special cases
\bea
q_{w0,1} + q_{w0,-1} & = & 
\frac{\ep_w}{\ep_\ell} q_0 
\label {41}\\
q_{w0,1} - q_{w0,-1} & = & q_0 \\
q_0 + q_{-1,-1} & = & 
\frac{\ep_\ell}{\ep_c} q_{c0,1} \\ 
q_0 - q_{-1,-1} & = & q_{c0,1}
\label {45} \\
q_{-1,1} & = & \frac{\ep_\ell}{\ep_c}
q_{c0,-1} \label {44} \\
q_{-1,1} & = & \mbox{} - q_{c0,-1}
\label {46}
\eea
the last two of which
imply that
\beq
q_{-1,1} = q_{c0,-1} = 0.
\label {zeros}
\eeq
The four equations 
(\ref{410}--\ref{440}) tell us that
for \(n>0\) and \(s=\pm1\)
\bea
q_{n,s} & = & -
\frac{\ep_w - \ep_\ell}{2\ep_w} q_{w-n,-s}
\label {51}\\
q_{-n,-s} & = & 
\frac{\ep_w + \ep_\ell}{2\ep_w} q_{w-n,-s}\\
q_{n,s} & = & 
\frac{\ep_c + \ep_\ell}{2\ep_c} q_{c n,s}\\
q_{-(n+1),-s} & = & -
\frac{\ep_c - \ep_\ell}{2\ep_c} q_{c n,s}
\label {54}
\eea
from which we can infer that for \(n > 0\)
\beq
q_{n,s} = -p q_{-n,-s} 
\label {qns=-pq-n-s}
\eeq 
and that for \(n>1\) and \(s=\pm1\)
\bea
q_{n,s} & = & (pp')^{n-1} q_{1,s} 
\label {recurrence 1}\\
q_{-n,s} & = & 
- p^{n-2}p^{\prime n-1} q_{1,-s} \\ 
q_{cn,s} & = & 
(1+p') (pp')^{n-1} q_{1,s} \\
q_{w-n,s} & = & 
-(1+p) p^{n-2} p^{\prime n-1} q_{1,-s} .
\label {recurrence 4}
\eea
The four relations (\ref{41}--\ref{44})
imply that
\bea
q_{w0,-1} & = & pq_{w0,1}
\label {61}\\
q_0 & = & (1-p) q_{w0,1} \\
q_{c0,1} & = & (1-p)(1+p') q_{w0,1} \\
q_{-1,-1} & = & - (1-p) p' q_{w0,1}.
 \label {64}
\eea
Gauss's law (\ref{VD=r}) applied
to a tiny sphere about the physical charge
\(q\) gives
\beq
q_{w0,1} = q.
\label {physical charge}
\eeq
This identification and 
the four equations (\ref{61}--\ref{64}) tell us that
\bea
q_{w0,-1} & = & pq
\label {71}\\
q_0 & = & (1-p) q \\
q_{c0,1} & = & (1-p)(1+p') q \\
q_{-1,-1} & = & - (1-p) p' q.
 \label {74}
\eea
Equations (\ref{qns=-pq-n-s}--\ref{74})
allow us to relate all the coefficients
for \(n > 0\) to \(q_{w0,1} = q\) 
and to \(q_{1,-1} = q_{-1,1} = 0\):
\bea
q_{n,1} & = & (pp')^n (1-p) q
\label {qn1=} \\
q_{n,-1} & = & (pp')^{n-1} q_{1,-1} = 0
\label {qn-1=} \\
q_{-n,1} & = & - p^{n-2} p'^{n-1} q_{1,-1} = 0
\label {q-ns=} \\
q_{-n,-1} & = & -p^{n-1} p^{\prime n} 
(1-p) q
\label {q-ns=} \\
q_{w-n,1} & = & - (1+p)
p^{n-2}p^{\prime n-1} q_{1,-1} = 0
\label {qw-n1=} \\
q_{w-n,-1} & = & - (1-p^2)
p^{n-1}p^{\prime n} q
\label {qw-n-1=} \\
q_{cn,1} & = & (1-p)(1+p')
(pp')^n q
\label {qcn1=} \\
q_{cn,-1} & = & (1+p')
(pp')^{n-1} q_{1,-1} = 0.
\label {qcn-1=}
\eea
\par
\par
The electric potential due to a charge \(q\)
in the extra-cellular environment
a distance \(h\) above a lipid bilayer 
of thickness \(t\) then is 
\bea
V_\ell(\rho,z) & = &
\frac{q}{4\pi\epsilon_0 \epsilon_{w\ell}}
\, \sum_{n=0}^\infty
(p p')^n \lt(
\frac{1}{\sqrt{\rho^2 + (z-2nt-h)^2}} \rt. \nn\\
& & \lt. \mbox{} 
- \frac{p'}{\sqrt{\rho^2 + (z+2(n+1)t+h)^2}}
\rt)
\label {Vellt2}
\eea
in the lipid bilayer.
That in the extra-cellular environment is
\bea
V_w(\rho,z) & = & 
\frac{q}{4\pi\epsilon_0 \epsilon_w}
\lt(\frac{1}{r}
+ \frac{p}{\sqrt{\rho^2 + (z+h)^2}} 
\rt. \label {Vexth2}\\
& & \mbox{} -  \lt. 
\frac{\ep_w \ep_\ell}{\ep_{w\ell}^2}
\sum_{n=1}^\infty
\frac{p^{n-1} p^{\prime n}}
{\sqrt{\rho^2 + (z + 2nt + h)^2}} \rt)
\nn
\eea
in which \(r\) is the distance from the
charge \(q\)\@.
Finally, the potential 
in the cytosol is 
\beq
V_c(\rho,z) = \frac{q \, \ep_\ell}
{4\pi\ep_0\ep_{w\ell}\ep_{\ell c}}
\! \sum_{n=0}^\infty
\frac{(p p')^n}{\sqrt{\rho^2 + (z-2nt-h)^2}}.
\label {Vct2}
\eeq
\section{Second Electrostatic Problem
\label{Second Electrostatic Problem}}

Here I approximate the electrostatic
potential \(V_\sigma(\bos{r}_k)\) 
within a disk of radius \(R\)
due to a uniform charge density
\(\sigma\) of phosphatidylserines (PSs)
outside the disk.
\par
The negative charges of the PSs 
are taken to lie on the interface between
the cytosol and the lipid bilayer.
The role of this potential \(V_\sigma\)
is only to keep the mutual repulsion 
of the PSs inside the disk from driving them
too much toward the perimeter of the disk.
So an exact expression
for \(V_\sigma\) is not needed.
Any formula for it will involve
an integral of \(\sigma\) over distances 
that run to infinity.
My approximation is to set the 
thickness \(t\) of the bilayer equal to zero.
In this limit, the effective potential
felt by a PS at \( \bos{r}_k \) is
\beq
V_\sigma(\bos{r}_k) = 
\sigma \int_R^\infty \!\! d\rho \int_0^{2\pi} \!\! d\phi
\,\, \frac{\rho}{q} \, V_{ws}(|\bos{\rho} - \bm{r_k}|)
\label {B 1}
\eeq
in which \(\bos{\rho} = \rho (\cos \phi,\sin\phi,0)\),
and \(V_{ws}(|\bos{\rho} - \bm{r_k}|)\) is the potential \(V_w(\rho,z)\)
of Eq.~(\ref{Vexth2}) for \(z = h = t = 0\) 
\beq
V_{ws}(|\bos{\rho} - \bm{r_k}|) =  
\frac{q}{4\pi\epsilon_0 \ep_{wc} |\bos{\rho} - \bm{r_k}|}
\label {Vws}
\eeq
where \( \ep_{wc} = (\ep_w + \ep_c)/2 \)\@.
With this approximation and with
\(\ep_w \approx \ep_c \approx 80\),
the potential (\ref{B 1}) is
\bea
V_\sigma(\bm{r_k}) & \approx &
\sigma \int_R^\infty \!\! d\rho \int_0^{2\pi} \!\! d\phi
\,\, \frac{\rho}
{4 \pi \ep_0 \ep_w |\bos{\rho} - \bm{r_k}|} 
\label {B 3}\\
 & = &
\frac{\sigma}{4 \pi \ep_0 \ep_w} 
\int_R^\infty \!\! d\rho \int_0^{2\pi} \!\! d\phi
\,\, \frac{\rho}
{\sqrt{\rho^2 + r_k^2 -2\rho r_k \cos \phi}}
\nn\\
& = & \frac{\sigma}{4 \pi \ep_0 \ep_w} 
\int_R^\infty \!\! d\rho \int_0^{2\pi} \!\! d\phi
\sum_{n=0}^\infty \lt( \frac{r_k}{\rho} \rt)^n \,
P_n(\cos\phi) .
\nn
\eea
The \(n = 0\) term in this sum 
is an infinite constant, which
we drop because it does not
affect the containment of the PSs
within the disk.  The remaining terms are
\bea
V_\sigma(\bm{r_k}) & \approx &
\frac{\sigma}{4 \pi \ep_0 \ep_w} 
\int_R^\infty \!\! d\rho 
\sum_{n=1}^\infty \lt( \frac{r_k}{\rho} \rt)^{2n} \,
2 \pi \lt[ {2n \choose n} 2^{-2n} \rt]^2 \nn\\
& = & \frac{\sigma R}{2 \ep_0 \ep_w} 
\sum_{n=1}^\infty \frac{1}{2n-1} 
\lt[ \frac{(2n)!}{(n!)^2 2^{2n}} \rt]^2  
\lt( \frac{r_k}{R} \rt)^{2n}.
\label {B 4}
\eea

\begin{acknowledgments}
I am grateful to Leonid Chernomordik 
for tips about electroporation, to
Gisela T\"{u}nnemann for sharing her data, 
to Sergio Hassan for advice about the
NaCl potential,
and to Pavel Jungwirth for advice on 
guanidinium groups,
to John Connor and Karlheinz Hilber for 
explaining the status of measurements
of the membrane potential of mouse myoblast cells,
to Paul Robbins for sending me some
of his images, to Jean Vance for
information about mammalian cells
deficient in the synthesis of 
phosphatidylserine,
and to James Thomas for 
several useful conversations.
Thanks also to H.~Berg, S.~Bezrukov, H.~Bryant, 
P.~Cahill, D.~Cromer, G.~Herling,
T.~Hess, S.~Koch,
V.~Madhok, M.~Malik, A.~Parsegian, 
B.~B. Rivers, K.~Thickman, and T.~Tolley
for useful conversations, and to  
K.~Dill, S.~Dowdy, S.~Henry, K.~Hilber, 
A.~Pasquinelli, 
B.~Salzberg, D.~Sergatskov, L.~Sillerud, 
A.~Strongin, R.~Tsien, J.~Vance,
and A.~Ziegler 
for helpful e-mail.
\end{acknowledgments}
\bibliography{bio,physics}
\end{document}